\documentclass[aps,prl,floatfix,12pt]{revtex4-2}
\usepackage{amsmath}
\usepackage{amssymb}
\usepackage{graphicx}
\usepackage{bm}
\usepackage{color}
\usepackage{times}
\usepackage{booktabs}
\usepackage{multirow}
\usepackage{diagbox}
\UseRawInputEncoding

\begin{document}

\title{Additive rule of real and reciprocal space topologies at disclinations}

\author{Qinghua He$^{1,2}$}
\author{Jinhua Sun$^{1}$}
\author{Hai-Yao Deng$^{3}$}
\email{dengh4@cardiff.ac.uk}
\author{Katsunori Wakabayashi$^{4,5}$}
\email{waka@kwansei.ac.jp}
\author{Feng Liu$^{1,2}$}
\email{Liufeng@nbu.edu.cn}

\affiliation{$^{1}$School of Physical Science and Technology, Ningbo
  University, Ningbo, 315-211, China}
\affiliation{$^{2}$Institute of High Pressure Physics, Ningbo
  University, Ningbo, 315-211, China}
\affiliation{$^{3}$School of Physics and Astronomy, Cardiff University, 5 The Parade, Cardiff CF24 3AA, Wales, United Kingdom}
\affiliation{$^{4}$Department of Nanotechnology for Sustainable Energy, School of Science and Technology, Kwansei Gakuin University, Gakuen 2-1, Sanda 669-1337, Japan}
\affiliation{$^{5}$Center for Spintronics Research Network (CSRN), Osaka University, Toyonaka 560-8531, Japan}

\begin{abstract}
Topological materials are renowned for their ability to harbor states localized at their peripheries, such as surfaces, edges, and corners. Accompanying these states, fractional charges appear on peripheral unit cells. 
Recently, topologically bound states and fractional charges at disclinations of crystalline defects have been theoretically predicted. This  so-called bulk-disclination correspondence has been experimentally confirmed in artificial crystalline structures, such as microwave-circuit arrays and photonic crystals. 
Here, we demonstrate an additive rule between the real-space topological
 invariant $\mathbf{s}$ (related to the Burgers vector $\mathbf{B}$) and the reciprocal-space
 topological invariant $\mathbf{p}$ (vectored Zak's phase of bulk wave
 functions). The bound states and fractional charges concur at a
 disclination center only if $\mathbf{s}+\mathbf{p}/2\pi$ is 
topologically nontrivial; otherwise, no bound state forms even if fractional charges are trapped. Besides the dissociation of
fractional charges from bound states, the additive rule also dictates the existence of half-bound states extending over only half of a sample and ultra-stable bound states protected by both real-space and reciprocal-space
topologies. Our results add another dimension to the ongoing study of topological matter and may germinate interesting applications.    
\end{abstract}

\maketitle
Topology studies the properties of a geometric or physical system under
continuous transformations in parameter spaces. Its application in
condensed matter physics has, in the past few years, renewed our
understanding of energy band structures of crystalline
systems~\cite{Hasan2010,Qi2011,Bansil2016}. As a cornerstone, the
so-called bulk-edge correspondence
principle~\cite{Hatsugai1993,Liang2006,Hwang2019,Wang2021} requires that 
robust quantum states appear at the boundaries of samples possessing
topologically nontrivial band
structures~\cite{Fujita1996,Delplace2011,Kane2013,Hafezi2013}. This
principle links the reciprocal-space topology (i.e., energy band structure) to
real-space profiles of quantum states. It provides a foundation for
potentially transformative applications in spintronics and other
practical
areas. Recently~\cite{Fu2011,Shiozaki2014,Benalcazar2017,Langbehn2017,Song2017,Qian2021,Tan2022},
the bulk-edge correspondence has been extended to higher-order
topological phases, culminating in the discovery of topologically
protected corner
states~\cite{Peterson2018,Imhof2018,Serra-Garcia2018,Ota2019,Xie2019,Xue2020}.
Applications such as laser cavity and quantum computation 
have been proposed based on these states~\cite{Hararieaar2018, Wu2020}. 

Unlike edge states, 
topological corner states usually appear as bound states in the continuum of bulk spectrum, 
which complicates their experimental detection~\cite{Hsu2016, Benalcazar2020, Cerjan2020, Liu2021a}. 
However, higher-order topological phases induce fractional charges and bound states at a disclination center of crystallographic defects, which has been experimentally observed recently~\cite{Benalcazar2019,Li2020,Peterson2021,Liu2021}.  
The correlation between the fractional charges carried by bound states
appearing at disclination centers and the reciprocal topological invariant
of bulk is framed as the bulk-disclination 
correspondence~\cite{Ruegg2013,Juan2014,Jeffrey2016,Roy2021,Geier2021}. 
Inspired
by such observations, here we focus on the correlation between
these anomalous bound states and the real-space topology of disclinations
(in contrast to the reciprocal-space topology of bulk). We find that the
real-space topology of disclinations and the reciprocal-space topology
of bulk are additive, in the sense that bound states form at
a disclination only if the sum between them is nontrivial. This additive rule can be
manifested in various ways, amongst which are the dissociation of the
appearance of fractional charges from the formation of bound states, the
existence of half-bound states extending over only half of a sample, and the
transition from bound states protected by real-space topology to those
by reciprocal-space topology, which we detail in what follows. Our
results bridge the classical real-space topologies of crystalline defects to the reciprocal-space
topologies of quantum wave functions and may well fertilize interesting physical
phenomena and applications.   

Being global crystallographic defects, 
disclinations cannot be removed by local operations
 \cite{Gopalakrishnan2013}. To construct a 
disclination, one may employ the Volterra method~\cite{Kleman2008}. An example is depicted
in Fig.~1\textbf{a}, whereby a sample is cut into a
few identical wedge portions, and one (marked in yellow) is removed to
form a disclination after gluing the remaining sections. According to
the homotopy theory, a disclination  
is characterized by two parameters $(\Omega, \mathbf{B})$. Here $\Omega$ is the Frank angle, whose magnitude is the wedge angle and whose sign indicates adding or removing a wedge, and $\mathbf{B}$ is the Burgers vector, which measures the lattice distortion induced by the defect~\cite{Azevedo1998,Alexander2012}. For square lattice that respects $C_4$ point group symmetry, $\Omega$ can only be a multiple of $\pi/2$ and the group of non-equivalent classes of $\mathbf{B}$ is isomorphic to the discrete group $Z_2$ and $Z_2 \otimes Z_2$ for $\Omega=\pm\pi/2$ and $\pm\pi$, respectively~\cite{Jeffrey2016}.

To exemplify the aforementioned additive rule between the real-space topology of disclination and the reciprocal-space topology of the wave function of bulk, we consider the two-dimensional (2D) Su-Schrieffer-Heeger (SSH)
model~\cite{Liu2017}, which is one of the typical models that admit topological corner states~\cite{Ota2019,Xie2019,Liu2019,Xu2021,Supplement}. A sample of 
of the 2D SSH model is depicted in Fig.~1\textbf{b}, where the unit cell
consists of four sub-lattices forming a square Bravais lattice. There are two types of hopping, namely the intra-cell hopping $\gamma$
and the inter-cell hopping $\gamma^\prime$. Depending on the ratio of
$|\gamma/\gamma^\prime|$, it settles in one of two topologically
distinct phases. For $|\gamma|<|\gamma^\prime|$ as in Fig.~1\textbf{b}, the lowest energy band
is inverted at $(\pi/a,0)$ and $(0,\pi/a)$ in the reciprocal-space (with
$a$ the lattice constant) and becomes topologically nontrivial
accompanying with corner states~\cite{Liu2017,Supplement}. The appearance of
topological corner states in the 2D SSH model is owing to the shift of
dimerized cells as displayed by the light magenta square in Fig.~1\textbf{b}, whose centers are related to the vectored Zak's phase $\mathbf{p} = (p_x,p_y)$
by a factor of
$\frac{a}{2\pi}$~\cite{Zak1989,Vanderbilt1993,Resta1994,Fang2012a}. Constrained
by the periodicity of Bravais lattice, $p_{x/y}$ is defined within $[0,2\pi)$
and becomes a quantization of $\pi$ when inversion symmetry is present, as
determined by the parity of the bulk wave function at $(0,0)$ and
$(\pi/a,0)/(0,\pi/a)$ in the reciprocal space. Upon shifting the center
of dimerized cells as well as Wannier states, the lowest energy band
accommodates less than one electron in the unit cells located at edges
and corners, known as the filling-anomaly that results in topological edge
and corner states carrying $1/2$ and $1/4$ fractional charges,
respectively.~\cite{Hwang2019,Supplement}.   

Figure 1\textbf{c} displays two distinct disclinations with
$\Omega=-\pi/2$ for the 2D SSH model, where the square represents the unit cell and the intra-cell and inter-cell hoppings are omitted. Depending on the Burgers vector $\mathbf{B}$ (red vectors in Fig.~1\textbf{c}), the disclinations of $\Omega=-\pi/2$ are classified into two topologically distinct types as labeled by $\mathbf{s}=(0,0)$ and $\mathbf{s}=(1/2,1/2)$, respectively. The relation between $\mathbf{B}$ and $\mathbf{s}$ is given as $\mathbf{s} =
\frac{1}{2}\left[(2\mathbf{B}) ~\text{mod} ~2\right]$, which forms a bijection to the homotopy group of $\mathbf{B}$ and thus is a real-space topological invariant. For a finite sample with full point-group symmetry, $\mathbf{s}$ can also be determined by counting the number of unit cells along the boundaries of
the sample, i.e., $\mathbf{s} = \frac{1}{2} [(\Gamma_x,\Gamma_y) ~
\text{mod}~2]$, where $\Gamma_x$ and $\Gamma_y$ denote the numbers of
unit cells on $x$- and $y$- boundaries, respectively.

Considering that the
removal or addition of the wedge part resolves the filling anomaly at the disclination center, we expect a concurrent action of the
real-space topological invariant $\mathbf{s}$ and the reciprocal
topological invariant $\mathbf{p}$, which we propose as an additive
rule between them. In TABLE~I, $\mathbf{s}$ is tabulated for all
possible values of $\Omega$ for the 2D SSH model~\cite{Omega}. The
integers inside TABLE~I are the numbers of bound states at the different
types of disclination centers for both trivial and nontrivial reciprocal
topologies. From TABLE~I, we see that even for the trivial reciprocal topology,
bound states exist as $\mathbf{s}+\mathbf{p}/2\pi$ is nontrivial, whereas for the nontrivial $\mathbf{p}$ bound state is missing if $\mathbf{s}+\mathbf{p}/2\pi$ is trivial. We define the net topology of real-space and reciprocal topologies as $\bm{\mathcal{P}}=2(\mathbf{s}+\mathbf{p}/2\pi) ~\text{mod} ~2$, and discuss three unique manifestations of the proposed additive rule in the follows, which embody the content in TABLE~I.  


The first manifestation is dissociation of fractional
charges from bound states. We demonstrate this
phenomenon for samples with $\left(-\frac{\pi}{2}\right)$-disclinations.
Figures 2\textbf{a}-\textbf{c} show the fractional charges and bound states for the 
$\left(-\frac{\pi}{2}\right)$-disclinations with three distinct 
additive conditions between the real and reciprocal topological invariants $\mathbf{s}$ and $\mathbf{p}$. 
In the left panels of Figs.~2\textbf{a}-\textbf{c}, the numerical datum of charge distribution 
for each unit cell are written.  
The bound states are indicated by the dark magenta shades (circles and triangles), and 
the fractional charges with $\pm 1/4$ are marked with the cyan
crescents. 
In the right panels of Figs.~2\textbf{a}-\textbf{c}, 
we have also displayed the numerical datum of eigenfunctions when electrons are mostly localized for the corresponding left samples at disclination centers.

As can be seen in the left panel of Fig.~2\textbf{a}, fractional
charges appear at the disclination center and the sample corners, but bound states are absent at the center (see also the right
panel of Fig.~2\textbf{a}) even with the nontrivial reciprocal topology $\mathbf{p}$. 
This result can be intuitively understood  using the dimerization
of sites as shown by lighter magenta squares in the left panel of
Fig.~2\textbf{a}. 
As explained earlier, the corner state accompanying $1/4$ fractional charge appears due to dimerized cells shifting from the original Bravais lattice and the resulting filling anomaly. However,
here in Fig.~2\textbf{a}, the filling anomaly at the disclination center that is supposed to be induced by nontrivial $\mathbf{p}$ is canceled out by the nontrivial real-space topological invariant $\mathbf{s}$. As a result, there is no fractionally filled dimerized cell isolated from the bulk states as indicated by the additive rule.
We note that the
fractional charge appears at the disclination center in Fig.~2\textbf{a}
simply because of the missing of a site in the central unit cell of the sample.

Figure~2\textbf{b} shows the disclination with trivial $\mathbf{s} =
(0,0)$ but non-trivial $\mathbf{p} = (\pi,\pi)$.  
Since the additive rule gives nontrivial $\bm{\mathcal{P}}$, both the bound
states and fractional charges simultaneously appear at the disclination center together with corner states as seen in Fig.~2\textbf{b}. 
Figure~2\textbf{c} shows a complementary example, where
the real-space topology is nontrivial, and the reciprocal space topology
is trivial. 
The additive rule gives nontrivial $\bm{\mathcal{P}}$. 
Thus, both the bound state and $1/4$ fractional charge appear at the center of
disclination without corner states as can be seen in Fig.~2\textbf{c}.
We shall note that the fractional
charge at the disclination center is further smeared out beyond the fractionally filled dimerized cell as seen in the left and right panels of Fig.~2\textbf{c}.

The second manifestation is the formation of half-bound states,
which decay on one side of the sample but extend over the other. 
The half-bound state exists if $s_x$ differs from $s_y$, i.e., with asymmetry
between the $x$ and $y$-directions. 
According to TABLE I, we need to
consider disclinations with $\Omega = -\pi$, and 
$(s_x, s_y)=(0, 1/2)$ or $(1/2,0)$.
The disclination of this class is shown in Figs.~3\textbf{a} and \textbf{b},  
where $\Omega=-\pi$ and $\mathbf{s} = (0,1/2)$, but for different
$\mathbf{p}$ ($\gamma/\gamma^\prime=1/3$ in \textbf{a} and $\gamma/\gamma^\prime=3$ in \textbf{b}). 
In this case, net topology $\bm{\mathcal{P}}$ is nontrivial for both
$\mathbf{p} = (0,0)$ and $\mathbf{p} = (\pi,\pi)$. 
Thus the half-bound states appear as shown in Figs.~3\textbf{a} and \textbf{b},  
where electrons decay in one direction but extend to the other direction. The decaying direction of half-bound state depends on which component of $\bm{\mathcal{P}}$ is nontrivial, i.e., nontrivial $\mathcal{P}_y$ yields a half-bound state decaying over the $x$-side. 
The formation of half-bound states is analogous to 
the formation of edge states due to the second-order topology.
In the 2D SSH model, if the systems have $p_xp_y = 0$ but $p_x+p_y \neq 0$, only edge
states exist but no corner state~\cite{Liu2019}. 
In the present case, this may be
paraphrased: For two-sided systems with $s_xs_y=0$ but
$s_x+s_y\neq0$, only a half-bound state exists but not a bound state. This
half-bound state can potentially control
wave-propagation using artificial crystalline structures such as
photonic crystals. These states are impervious of the system
size~\cite{Supplement}.

The third manifestation can be observed in any disclination with $\Omega
\ge \pi$ and $s_x\neq s_y$. 
Figure~4\textbf{a} shows a disclination with $\Omega=\pi$ and
$\mathbf{s} = (1/2,0)$. 
This disclination is formed by inserting two extra $\pi/2$ blocks into
the sample, which has three $x$-sides and three $y$-sides arranged alternatingly. Owing to the additional blocks of $\pi/2$, the wave function may decay in multiple directions (even when only the $x$-side is nontrivial), and bound state forms rather than the half-bound state. Furthermore, $\bm{\mathcal{P}}$ is
nontrivial regardless of $\mathbf{p}$ being trivial or
nontrivial. Hence, at the
disclination center, bound states form invariably for \textit{arbitrary parameters}. 
Figure 4\textbf{b} displays the energy spectrum for the disclination in
Fig.~4\textbf{a} with $\mathbf{p}=(0,0)$, where doubly degenerate bound
state emerges within the band gap. 
Interestingly, for $\mathbf{p}=(\pi,\pi)$, 
the bound state becomes a singlet with a symmetric wave function, as illustrated in Fig.~4\textbf{c}. 
This phenomenon reflects on the different topological
origins of the bound states. The doublet bound state originates from the real-space topology, which distinguishes the three non-equivalent $\pi/2$ blocks of $y$-side in the real space. On the other hand, the singlet bound state does not differentiate those blocks in real space owing to its reciprocal topological origin. The difference between doublet and singlet bound states reminds us the conjugation relation between real and reciprocal spaces. The bound states at such disclinations are ultra-stable and protected by both real-space and reciprocal-space topologies. They are useful for nano-scale photonic cavities. A full spectrum of parameter pumping for such the bound states is given in Supplement~\cite{Supplement}.

Finally, we remark on the number of bound states at a disclination as listed in TABLE I for $\mathbf{p} = (0,0)$. For $\Omega=-\pi$ and $\mathbf{s} = (1/2,1/2)$, there are two bound states with opposite energies due to chiral symmetry. For $\Omega=\pm \pi/2$ and $\mathbf{s} = (1/2,1/2)$, there is a pair of degenerate bound states due to the extra nontrivial $\pi/2$ block. For $\Omega=\pi$, there are two pairs of degenerate states with opposite energies due to chiral symmetry and the additional blocks.

To summarize, we have demonstrated an additive rule between the
real-space topology and the reciprocal-space topology at typical
crystallographic defects, namely the disclinations. The real-space
topology is characterized by the parity of the Burgers vector while the
reciprocal-space topology by the vectored Zak's phase.  
We demonstrate three unique phenomena due to the additive rule: the
dissociation of the bound state formation and the appearance of
fractional charges, the existence of half-bound states, and ultra-stable
bound states protected by real- and reciprocal-space topologies. Our
results shed further insight into crystalline topology and may usher in
novel applications.   

This work is supported by the Research Starting Funding of Ningbo
University, NSFC Grant No. 12074205, and NSFZP Grant No. LQ21A040004. K.W. acknowledges the financial support by JSPS KAKENHI Grant No. JP18H01154, and JST CREST Grant No. JPMJCR19T1.


\bibliography{references}

\begin{thebibliography}{52}
\bibitem [{Ome()}]{Omega}%
  \BibitemOpen
  \href@noop {} {}\bibinfo {note} {For $\Omega$ larger than $\pi$, it repeats
  the results of $\Omega=\pi/2$ and $\Omega=\pi$ periodically.}\BibitemShut
  {Stop}%
\end{thebibliography}
\clearpage

%

\begin{table*}
\begin{tabular*}{0.8\textwidth}{@{\extracolsep{\fill}}c|cccc|cccc}
&\multicolumn{4}{c|}{$\mathbf{p}=(0,0)$}
&\multicolumn{4}{c}{$\mathbf{p}=(\pi,\pi)$} 
\\\hline
\diagbox[width=4.2em]{$\mathbf{S}$}{$\Omega$} & 
          $-\pi$ & 
          $-\frac{\pi}{2}$ & 
          $\frac{\pi}{2}$  & 
          $\pi$ &
          $-\pi$ & 
          $-\frac{\pi}{2}$ & 
          $\frac{\pi}{2}$  & 
          $\pi$
\\\hline\hline
$(1/2,1/2)$ & 2 & 2 & 2 & 4 & 0 & 0 & 0 & 0 \\
$(0,1/2)$   & 0.5 & --- & --- & 4 & 0.5 & --- & --- & 2 \\
$(1/2,0)$   & 0.5 & --- & --- & 4 & 0.5 & --- & --- & 2 \\
$(0,0)$     & 0 & 0 & 0 & 0 & 2 & 2 & 2 & 4 \\ 
\hline
\end{tabular*}
\caption{
\textbf{Number of bound states for different disclination types and reciprocal topologies.} 
The disclination is characterized by the real space topological
invariant $\mathbf{s}$ and the Frank angle $\Omega$. $\Omega$ takes value of
$-\pi$, $-\frac{\pi}{2}$, $\frac{\pi}{2}$ and $\pi$. 
The reciprocal topological invariant, namely the vectored Zak phase $\mathbf{p}$, is $(0,0)$ for the trivial topological phase and $(\pi,\pi)$ for the nontrivial topological phase.
``0.5'' indicates a half-bound
 mode. ``---'' indicates such the type of disclination does not exist.
 } 
\end{table*}

\clearpage

\begin{figure}[!htp]
\begin{center}
\includegraphics[clip=true,width=1.0\columnwidth]{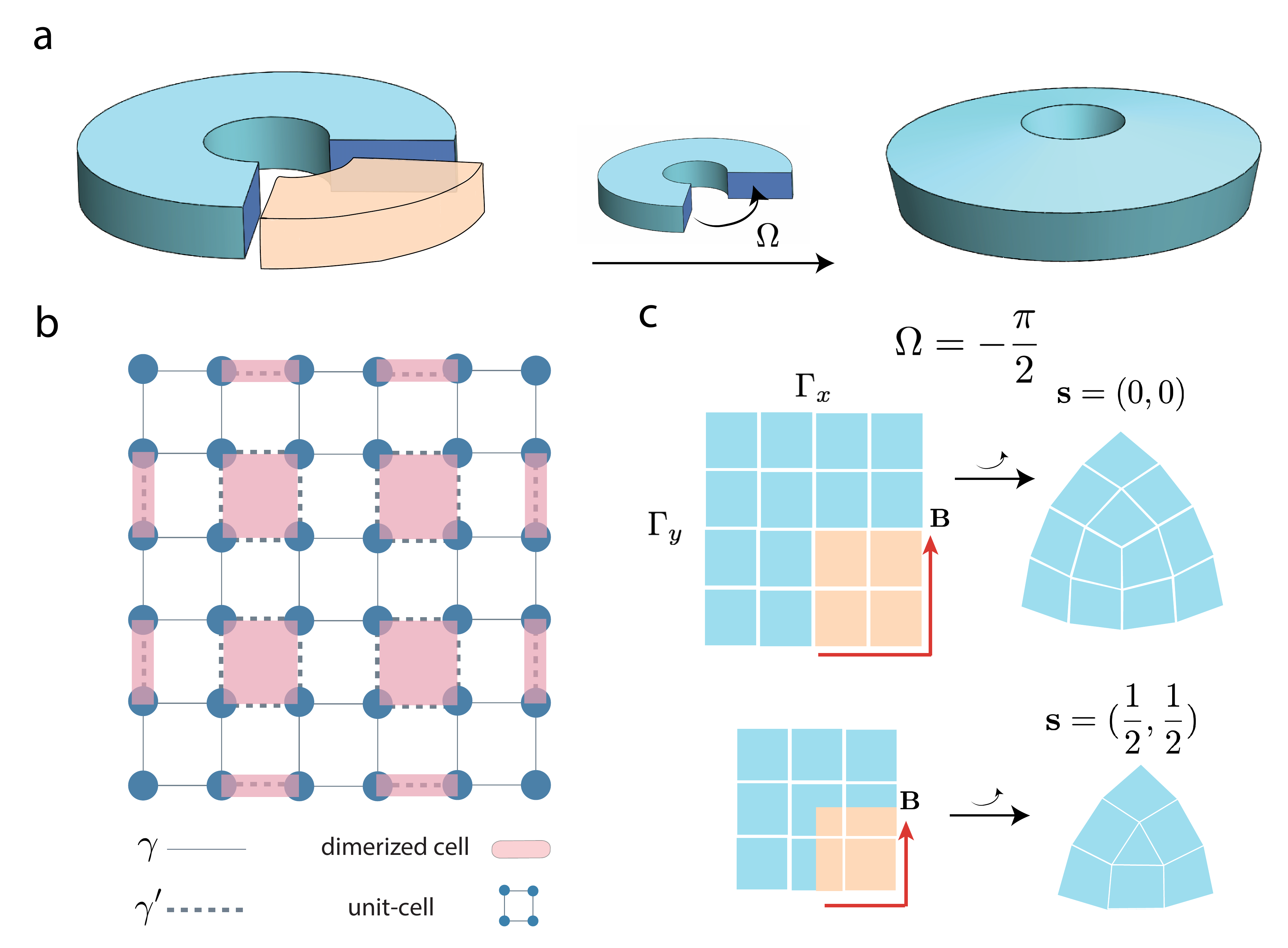}
\caption{
  \textbf{Construction and characteristic of a disclination.}
  ~\textbf{a}.~Schematic of Volterra process for constructing a
 disclination. A wedge part spanning angle $|\Omega|$ is cut off from a
 symmetric sample, and the remaining sections are glued without any
 lattice mismatch. The wedge center is located at the point of rotation
 symmetry of the sample. The resulting disclination has negative Frank
 angle $\Omega = - |\Omega|$. Alternatively, one may insert an extra wedge instead of removing the wedge, resulting in a disclination with
 positive $\Omega = |\Omega|$. \textbf{b}.~ Sample of the 2D SSH model in the case of $|\gamma|<|\gamma^\prime|$ that respects $C_4$ point
 group symmetry, where solid/dashed line indicates the intra/inter-cell
 hopping of strength $\gamma$/$\gamma^\prime$, and square/shade indicates the unit/dimerized cell. \textbf{c}.~Two types of disclinations with $\Omega=-\pi/2$ allowed for samples with $C_4$-point group symmetry characterized by $\mathbf{s}$. Each square represents a unit cell, and the lighter ones are the wedges being removed. $\mathbf{s}$ is determined by the parity of the numbers of
 unit cells on the x- and y- boundaries as $\mathbf{s} = \frac{1}{2}
 [(\Gamma_x,\Gamma_y) ~\text{mod}~2]$, which forms a bijection of the
 homotopy group of the Burgers vector $\mathbf{B}$. } 
\end{center}
\end{figure}

\begin{figure}[!htp]
\begin{center}
\leavevmode
\includegraphics[clip=true,width=0.78\columnwidth]{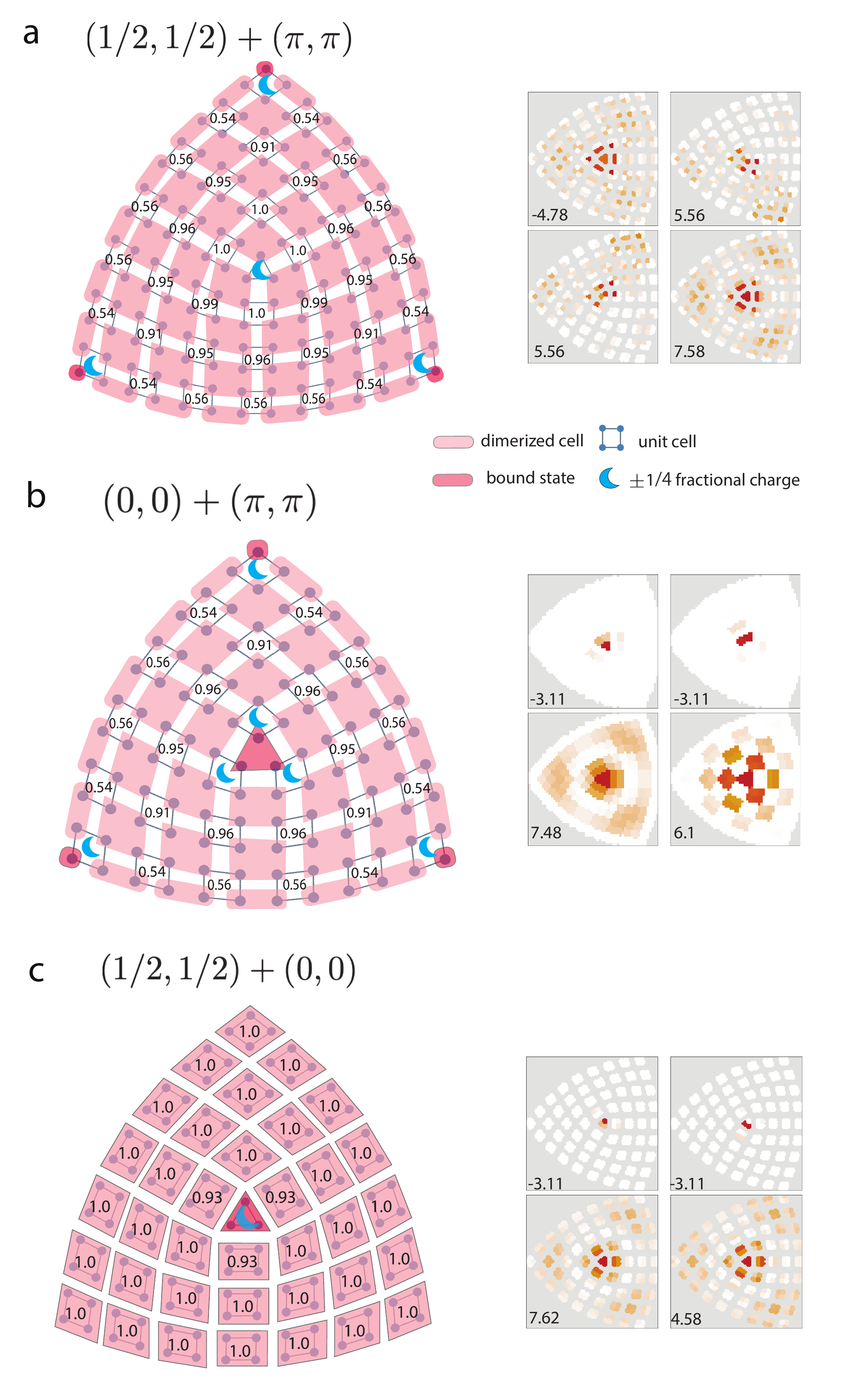}
\end{center}
\end{figure}

\begin{figure}[!htp]
\leavevmode
\caption{\textbf{Fractional charge and bound state dissociation for $(-\frac{\pi}{2})$-disclinations.} In left panels, a unit cell consists of 4 sites joined by thin lines, and a dimerized cell consists of sites in the same shade. Unit cells carrying $1/4$ fractional charge are each indicated by an crescent, and the numerical results of charge distribution are indicated by number in each unit cell. The fractional charge is calculated with four of the lowest energy band filled. Right panels are the most  concentrating 4 eigen states at the disclination center with their eigen energies indicated on the left corners.~\textbf{a}.~$\mathbf{s} = (1/2,1/2)$ and $\mathbf{p} = (\pi, \pi)$, and $\bm{\mathcal{P}}$, which denotes the parity of $\mathbf{s} + \frac{\mathbf{p}}{2\pi}$, is trivial. Hence, while fractional charge appears at the disclination center, there is no bound state therein.~\textbf{b}.~~$\mathbf{s} = (0,0)$ and $\mathbf{p} = (\pi, \pi)$, giving a nontrivial $\bm{\mathcal{P}}$. As a result, bound states appear at the disclination center along with fractional charges.~\textbf{c}.~ $\mathbf{s} = (\frac{1}{2},\frac{1}{2})$ and $\mathbf{p} = (0,0)$, and hence $\bm{\mathcal{P}}$ is also nontrivial, leading to both bound state and fractional charge at the disclination center like in \textbf{b}. }
\end{figure}
\clearpage

\begin{figure}[!htp]
\begin{center}
\leavevmode
\includegraphics[clip=true,width=1.0\columnwidth]{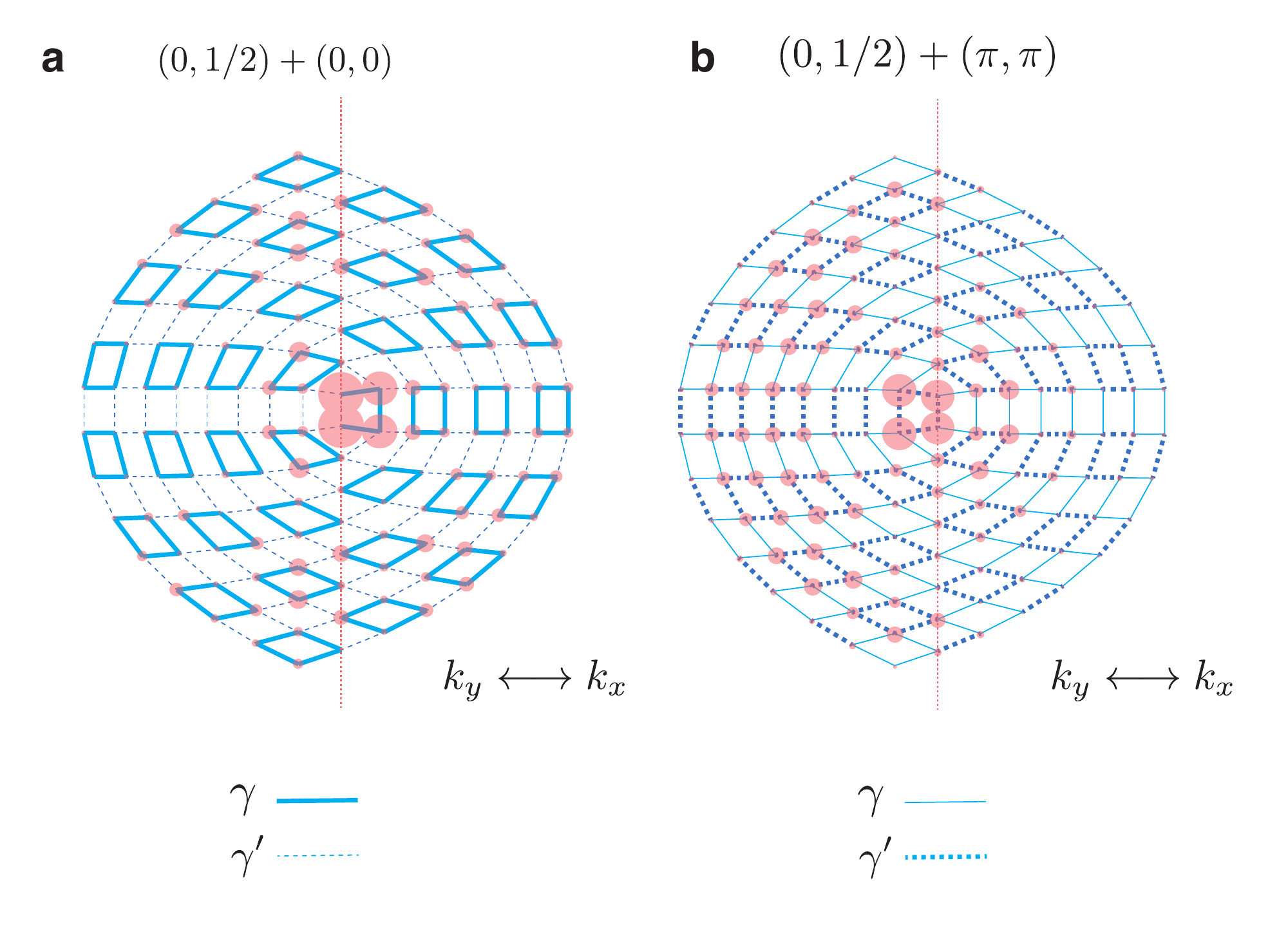}
\caption{\textbf{Existence of half-bound states. }The disclinations have $\Omega = -\pi$ and $\mathbf{s} = (0,1/2)$.~\textbf{a}.~$\mathbf{p} = (0,0)$, the state decays on the $x$-side but extends over the $y$-side.~\textbf{b}.~$\mathbf{p} = (\pi,\pi)$, it decays on the $y$-side but extends over the $x$-side on the other hand. A dashed line passes through the center of the disclination, which divides the sample into $x$- (perpendicular to $k_y$ direction of the reciprocal space) and $y$- (perpendicular to $k_x$ direction of the reciprocal space) parts. 
}
\end{center}
\end{figure}

\begin{figure}[!htp]
\begin{center}
\leavevmode
\includegraphics[clip=true,width=1.0\columnwidth]{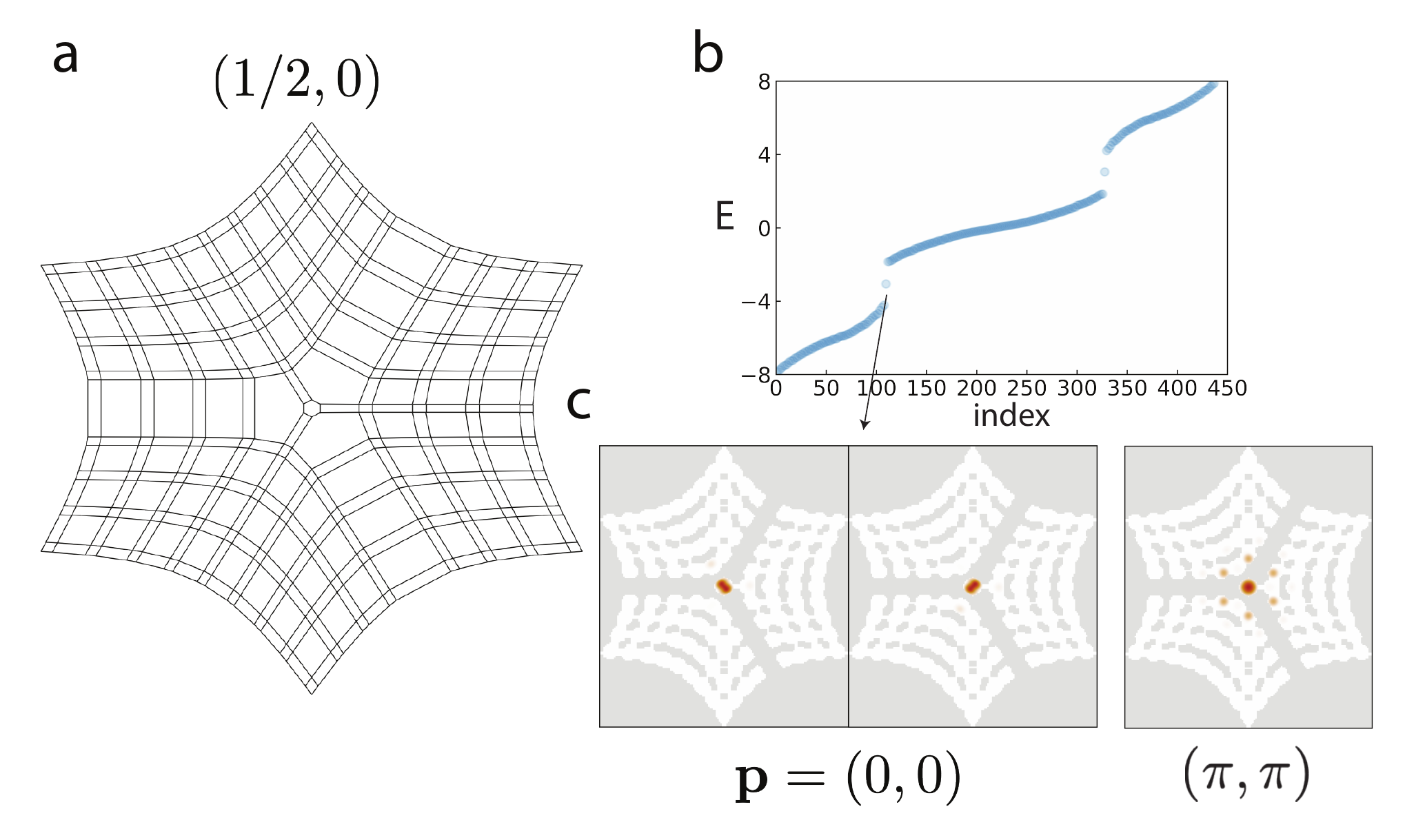}
\caption{\textbf{Real-space topology protected bound states and reciprocal-space topology protected ones.}~\textbf{a}.~The disclination has $\Omega = \pi$ and $\mathbf{s} = (1/2,0)$, and hence a nontrivial $\mathbf{s}+\mathbf{p}/2\pi$ irrespective of $\mathbf{p}$.  
\textbf{b} - \textbf{c}, Topologically stable bound states invariably emerge at the disclination center. The energy levels are displayed for $\mathbf{p} = (0,0)$ in \textbf{b}, where doubly degenerate in-gap bound states appear. However, for $\mathbf{p} = (\pi,\pi)$, only a single bound state of symmetric distribution appears inside the bulk band gaps, as shown in \textbf{c}.
}
\end{center}
\end{figure}
\clearpage

\includegraphics[scale=0.85,page=1]{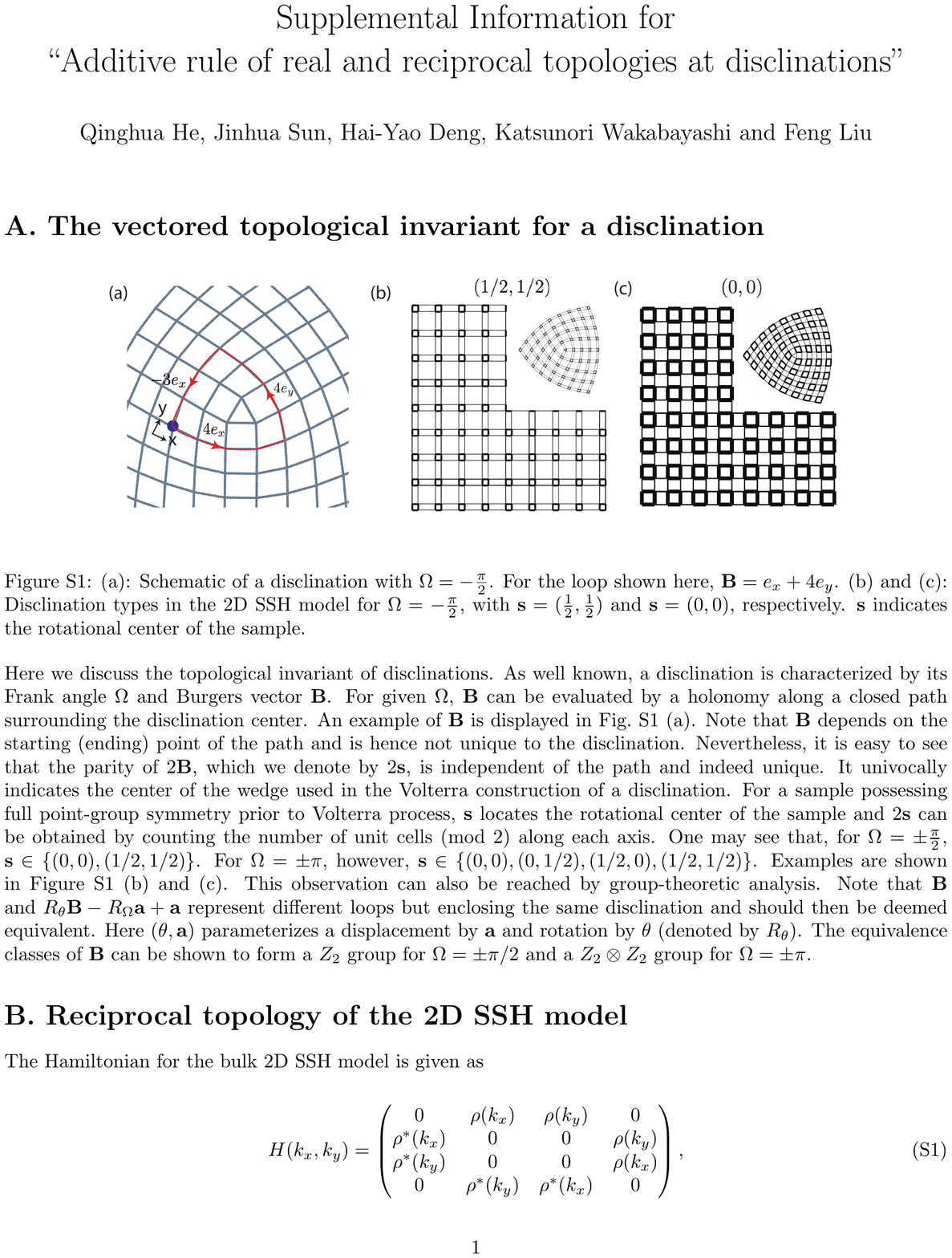}
\clearpage
\includegraphics[scale=0.85,page=2]{Supplementary.pdf}
\clearpage
\includegraphics[scale=0.85,page=3]{Supplementary.pdf}
\clearpage
\includegraphics[scale=0.85,page=4]{Supplementary.pdf}

\end{document}